\documentclass[twocolumn,floatfix]{aastex62}

\usepackage{savesym}
\savesymbol{tablenum}
\usepackage{siunitx}
\usepackage{bm}
\usepackage{mathtools}
\usepackage{cprotect}
\usepackage{amsmath}

\begin{document}

\title{Spectral Signatures of Population III and Envelope-stripped Stars in Galaxies at the Epoch of Reionization}

\correspondingauthor{Elizabeth Berzin}
\email{eberzin@princeton.edu}

\author{Elizabeth Berzin}
\affiliation{Department of Astrophysical Sciences, Princeton University, Peyton Hall, 4 Ivy Lane, Princeton NJ 08544, USA}

\author{Amy Secunda}
\affiliation{Department of Astrophysical Sciences, Princeton University, Peyton Hall, 4 Ivy Lane, Princeton NJ 08544, USA}

\author{Renyue Cen}
\affiliation{Department of Astrophysical Sciences, Princeton University, Peyton Hall, 4 Ivy Lane, Princeton NJ 08544, USA}

\author{Alexander Menegas}
\affiliation{Department of Astrophysical Sciences, Princeton University, Peyton Hall, 4 Ivy Lane, Princeton NJ 08544, USA}

\author{Ylva G\"{o}tberg} \thanks{Hubble Fellow} 
\affiliation{The Observatories of the Carnegie Institution for Science, 813 Santa Barbara Street, CA 91101 Pasadena, USA}

\begin{abstract}
While most simulations of the epoch of reionization have focused on single-stellar populations in star-forming dwarf galaxies, products of binary evolution are expected to significantly contribute to emissions of hydrogen-ionizing photons. Among these products are stripped stars (or helium stars), which have their envelopes stripped from interactions with binary companions, leaving an exposed helium core. Previous work has suggested these stripped stars can dominate the LyC photon output of high-redshift low luminosity galaxies post-starburst. Other sources of hard radiation in the early universe include zero-metallicity Population III stars, which may have similar SED properties to galaxies with radiation dominated by stripped star emissions. Here, we use four metrics (the power-law exponent over wavelength intervals 240-500 $\text{\AA}$, 600-900 $\text{\AA}$, and 1200-2000 $\text{\AA}$, and the ratio of total luminosity in FUV wavelengths to LyC wavelengths) to compare the SEDs of simulated galaxies with only single-stellar evolution, galaxies containing stripped stars, and galaxies containing Population III stars, with four different IMFs. We find that stripped stars significantly alter SEDs in the LyC range of galaxies at the epoch of reionization. SEDs in galaxies with stripped stars have lower power-law indices in the LyC range and lower FUV to LyC luminosity ratios. These differences in SEDs are present at all considered luminosities ($M_{UV} > -15$, AB system), and are most pronounced for lower luminosity galaxies. Intrinsic SEDs as well as those with ISM absorption of galaxies with stripped stars and Pop III stars are found to be distinct for all tested Pop III IMFs.
\end{abstract}

\section{Introduction} \label{sec:intro}

The James Webb Space Telescope (JWST) is expected to transform our understanding of high-redshift galaxies, and test our predictions for the epoch of reionization. Recently, galaxy simulations aimed at improving our understanding of the epoch of reionization have included stars that interact in binaries \citep{Ma_2016, Rosdahl_2018, Secunda2020}. Although binary fractions in low-metallicity environments like those of the early universe have yet to be measured directly, observations of massive stars in the Milky Way and Large Magellanic Cloud suggest binary interactions occur for approximately 70\% of young, massive stars in local environments \citep{Kobulnicky2007, Mason_2009, Sana2012, Almeida2017}. Through mass-transfer and coalescence, binary interactions can produce ionizing sources, such as high-mass stars or envelope-stripped helium stars, that emit ionizing photons tens to hundreds of Myr after starburst \citep{VanBever1999, 2016MNRAS.456..485S, Gotberg2019}. 

Because Lyman Continuum (LyC) photons from these products of binary interactions are ``delayed,'' the presence of binaries increases the total LyC emission in older stellar populations where there are no remaining O/B stars \citep{Eldridge2017, Secunda2020}. In addition, simulations that include ionizing photons from binary products show these photons are more likely to escape their host galaxy, as the delay provides more time for feedback from massive stars to remove the surrounding gas from the birth cloud, which normally traps LyC radiation \citep{KimmCen_2014, 2009Wise, Secunda2020, Ma_2016, Rosdahl_2018}. This increase in escaping LyC radiation that results from the presence of binary products suggests that the observed spectral energy distributions (SEDs) of galaxies that are 10-100 Myr post-starburst may look significantly different, primarily at ionizing wavelengths, than one would expect if no binary evolution occurred. 

Stripped stars are one such product of binary evolution, and are particularly interesting
because they have very high effective surface temperature and are copious LyC emitters \citep[see e.g.,][]{Gotberg2017}.
Stripped stars are formed when their envelopes are stripped from interaction with binary companions \citep{Kippenhahn67, Pols1994}, and emit significant ionizing radiation in stellar populations older than 10~Myr \citep{Gotberg2019}. \cite{Secunda2020} simulated high-redshift, dwarf galaxies with virial masses ($M_{vir}$) ranging from $10^8 - 10^{10.5}~M_{\odot}$, and found that the median rate of escaping photons of lower mass halos ($M_{vir} < 10^9\ M_{\odot}$) increases by a factor of as much as 200 when stripped stars are included. This result suggests that LyC emission from these low-mass halos is almost exclusively dominated by binary products. Despite the fact that a significant portion of massive stars undergo binary interactions to produce stripped stars, and the importance of stripped stars in the process of reionization \citep{Ma_2016, Rosdahl_2018, Secunda2020}, only a handful of stripped stars with main sequence companions have been observationally confirmed \citep[e.g.,][]{Groh2008, Wang2018}. If detected by JWST, low-mass galaxies dominated by stripped-star ionizing emission may provide an ideal laboratory for studying binary-interactions as a function of redshift.  

Other potential sources of ionizing radiation during the epoch of reionization are Population III (Pop III) stars. Pop III stars are first generation, zero-metallicity stars that were proposed early on by \cite{1978Natur.275...35R} in the context of the cosmic microwave background, and by \cite{1978MNRAS.183..341W} and \cite{1984ApJ...277..445C} to explain missing mass in galaxy clusters. Pop III stars are thought to have smaller radii, higher masses (up to several hundred solar masses), higher temperatures $(\sim 10^5\ \text{K})$, and a characteristically hard ionizing spectrum compared to typical massive stars \citep{Tumlinson_2000,Schaerer2002}. While it is unlikely that JWST will be able to probe isolated Pop III stars \citep{Rydberg_2013}, Pop III galaxies with stellar masses as low as $10^5\ M_\odot$ may be detectable \citep{Zackrisson_2011}. Because both stripped stars and Pop III stars are characterized by significant ionizing emission, high-redshift galaxies with stripped stars may masquerade as Pop III stars or vice versa. While galaxies containing Pop III stars may lack dust and have different gas-richness relative to galaxies with stripped stars, much is still unknown about the nature of Pop III galaxies. Thus, understanding the characteristics of SEDs of galaxies dominated by stripped stars is crucial to how we can distinguish between these two stellar populations.

Spectral shape in the ionizing regime is a sensitive probe for revealing what source is responsible for emission of ionizing photons \citep[see e.g.,][]{Gotberg2019}. Because the astrophysical objects mentioned here produce ionizing photons under different physical conditions, each of them are predicted to have a characteristic spectral hardness in the ionizing regime.

Apart from the most massive stars \citep{2002MNRAS.337.1309S, 2016MNRAS.458..624C} stripped stars, and Pop III stars, other sources of ionizing emission include accreting white dwarfs \citep{2015MNRAS.453.3024C}, X-ray binaries \citep{2019A&A...622L..10S, 2020MNRAS.494..941S}, and post-AGB stars \citep{2019AJ....158....2B}. Among these, stripped stars emit hydrogen-ionizing photons at the highest rates in stellar populations older than $\sim 10$ Myr. However, the accreting compact objects are predicted to have harder ionizing radiation and have therefore been considered to contribute substantially to the emission rate of He$^+$ ionizing photons \citep{2013MNRAS.432.1640W, 2020MNRAS.494..941S}. Chemical mixing and mass loss induced by moderate to extreme stellar rotation \citep{1987A&A...178..159M, 2005A&A...443..643Y} could also increase the emission rate of ionizing photons from a stellar population substantially \citep{2009A&A...497..243D, 2012ApJ...751...67L, 2015A&A...581A..15S, 2019A&A...623A...8K}. However, although circumstantial evidence have been claimed, direct proof that rotational mixing is efficient is still missing (see e.g., \citealt{2009A&A...495..257M, 2015A&A...581A..21H, 2018A&A...611A..75S, 2019A&A...627A.151S,2019ApJ...880..115A}). In this paper we choose to focus on and compare ionizing emission from massive stars, stripped stars and Pop III stars, only.

Recent observations have revealed remarkably hard ionizing spectra of stellar populations both in the local \citep{2019ApJ...878L...3B} and distant Universe \citep{2019A&A...624A..89N, 2020A&A...636A..47S}. No current spectral synthesis model produces a significant amount of sufficiently hard ionizing photons to explain the observed spectra, which highlights the urgency of better understanding sources of ionizing emission. Carefully characterizing the individual effects various ionizing sources have on the SEDs of galaxies is a necessary first step. These can then be used to produce observable predictions such as nebular line strengths of ionized elements \citep[cf.][]{2014MNRAS.444.3466S, 2018MNRAS.477..904X}.

In this paper, we seek to characterize the effects of stripped stars on the SEDs of galaxies at the epoch of reionization, and compare these results to predictions for Pop III stars. In \S \ref{sec:Method} we briefly describe the cosmological simulations used and the method for constructing the SEDs of galaxies. In \S \ref{sec:res} we discuss the SEDs of galaxies without accounting for absorption and in \S \ref{sec:abs} we discuss the SEDs of galaxies accounting for absorption by intervening gas and dust along randomly chosen sightlines. We summarize our results and conclude in \S \ref{sec:conclusion}.

\section{Methods}
\label{sec:Method}

\subsection{Cosmological Simulations}

We use output from the cosmological hydrodynamical simulations of \cite{KimmCen_2014}, performed using the \verb"RAMSES" Eulerian adaptive mesh refinement (AMR) code \citep{Teyssier2002}. \cite{KimmCen_2014} use the \verb"MUSIC" software \citep{Hahn_Abel_2011} to generate initial conditions with the following cosmological parameters from WMAP7 \citep{Komatsu_2011}:  ($\Omega_m$, $\Omega_\lambda$, $\Omega_b$, h, $\sigma_8$, $n_s$ = 0.272, 0.728, 0.045, 0.702, 0.82, 0.96). We expect that our results are only weakly dependent on these assumed cosmological parameters.

\cite{KimmCen_2014} begin with dark matter-only simulations of $256^3$ particles using a large volume of $(25\ \text{Mpc} h^{-1})^3$, in order to include the effects of the large-scale tidal field. They then zoom in on a $3.8 \times 4.8 \times 9.6$ Mpc box (comoving) and employ a series of refinements to achieve a dark matter mass resolution of $1.6 \times 10^5 M_\odot$. Further refinement, optimized to resolve the structure of the ISM, results in a minimum physical cell size of 4.2 pc, and a stellar mass resolution of approximately 49 $M_\odot$. Radiation from each star particle represents a full stellar population, rescaled for mass. These simulations include star formation, radiative cooling \citep{Sutherland1993, Rosen1995}, thermal stellar winds, feedback from supernova explosions, and photoionization by stellar radiation. For more details, see \cite{KimmCen_2014}.

\subsection{Stellar Populations}

In this section we describe the three stellar populations models used in this paper: a population accounting for only single stars, a population accounting for single stars and the creation of stripped stars, and a population consisting of only zero-metallicity Pop III stars. We also describe two versions of modelling for each population: one accounting for the presence of gas and dust and one not accounting for the presence of gas and dust. 

\subsubsection{Single Stars and Stripped Stars}
\label{sec:single_stripped_stars}

To model ionizing emission from single-stellar populations, we use Starburst99 \citep{Starburst99}, with stellar evolution models from Padova \citep{Bertelli_1993, Bertelli_1994, Marigo_2008}, atmospheric models from CMFGEN \citep{Hellier_1998} and WM-Basic \citep{Pauldrach_2001}, and a Kroupa initial mass function \citep{Kroupa2001} from 0.1 $M_\odot$ to 100 $M_\odot$. The metallicities available in Starburst99 are 0.0004, 0.004,  0.008,  0.02, and  0.05. 

We use the models of \cite{Gotberg2019} to simulate the emission of stripped stars. These are based on binary stellar evolution models created with MESA \citep{Paxton2011,Paxton2013,Paxton2015} and spectral models made with CMFGEN \citep{Hellier_1998}. For the predictions of a population, \cite{Gotberg2019} assume the same \cite{Kroupa2001} IMF as in the Starburst99 model and the mass dependent initial binary fraction synthesized by \cite{Moe_2017}. They assume the mass ratios, $q \equiv M_{2,init}/M_{1,init}$, where $M_{1,init}$ is the initial mass of the donor star and $M_{2,init}$ is the initial mass of the accretor star, are uniformly distributed between 0.1 and 1. They also assume the orbital periods are uniformly distributed in log-space following \cite{Opik1924} for $M_{1,init} < 15\ M_\odot$ and \cite{Sana2012} for $M_{1,init} > 15\ M_\odot$. The metallicities used are 0.0002, 0.002, 0.006, and 0.014. The ionizing emission from a stellar population containing stars stripped of their hydrogen-rich envelopes via binary interaction is modeled by combining the Starburst99 models with those of \cite{Gotberg2019}. Although Starburst99 is made for single stars, it can be used to represent radiation from binaries prior to interaction when their evolution is effectively single. For more details, see \cite{Gotberg2019}. 

\begin{figure}
\centering
\includegraphics[width=0.5\textwidth]{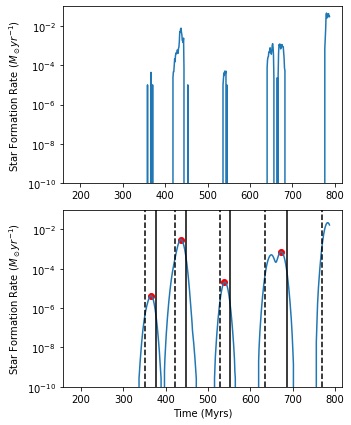}
\cprotect\caption{\textbf{Top:} Star formation history for a single galaxy (with stellar mass $4.27 \times 10^5\ M_\odot$ at $z = 7$) from the simulations of \cite{KimmCen_2014}. \textbf{Bottom:} Smoothed star formation history for a single galaxy from the simulations of \cite{KimmCen_2014}, corresponding to the original star formation history in the top panel. Maxima of relevant star-formation peaks are identified by red points, start-times of star formation peaks are identified by vertical dashed black lines, and end-times are identified by vertical solid black lines. The start and end time of a star formation peak is defined as the time at which star formation is a tenth of its maximum.}
\label{fig:sf_history}
\end{figure}

We interpolate between the different ages and metallicities of these stellar population models to generate SEDs for 586 galaxies from a snapshot of the \cite{KimmCen_2014} simulation at $z=7$ based on the ages, masses, and metallicities of each galaxy's star particles as output from the simulation. When calculating intrinsic SEDs (not accounting for absorption by intervening gas and dust), we recreate the star-formation history of each galaxy in order to construct multiple SEDs for the galaxy at various time points. To identify peaks of star-formation, we smooth the star-formation history of each galaxy, using a Gaussian kernel with a standard deviation of 5 Myr. An example of original and smoothed star formation histories for one of these galaxies is shown in the bottom panel of Figure \ref{fig:sf_history}, with the maxima of star-formation peaks identified by red points, start-times of star formation peaks identified by vertical dashed black lines, and end-times identified by vertical solid black lines. The start and end time of a star formation peak is defined as the time at which star-formation is 10\% of its maximum height. This smoothing is used only for the identification of start and end times of star formation peaks. To construct SEDs for simulated galaxies, we use the original, unsmoothed star-formation histories, as plotted in the top panel of Figure \ref{fig:sf_history}. SEDs are constructed for every Myr following each star formation peak, up to either 100~Myr or the beginning of the next star-formation peak.

When accounting for absorption by hydrogen, helium, and dust, the resulting luminosity at a particular frequency, $\nu$, and line of sight, $\Omega$, is given by:
\begin{equation}
\begin{split}
    L_{esc}(\nu, \Omega)  &=  L(\nu) \exp\biggl[ -\sigma_{\text{HI}}(\nu)N_{\text{HI}}(\Omega) \\ &
    - \sigma_{\text{HeI}}(\nu)N_{\text{HeI}}(\Omega) -\sigma_{\text{HeII}}(\nu)N_{\text{HeII}}(\Omega) \\ &   - k_{ext}\Sigma(\Omega) \biggr]
\end{split}
\end{equation}

where $ L(\nu)$ is the luminosity before absorption; $N_{\text{HI}}(\Omega)$, $N_{\text{HeI}}(\Omega)$, and $N_{\text{HeII}}(\Omega)$ are the neutral hydrogen, neutral helium, and singly-ionized helium column densities along the line of sight (until the edge of the simulation box), computed based on the gas distribution output from the \verb"RAMSES" simulation; $\sigma_{\text{HI}}(\nu)$, $\sigma_{\text{HeI}}(\nu)$, and $\sigma_{\text{HeII}}(\nu)$ are neutral hydrogen, neutral helium, and singly-ionized helium cross-sections calculated according to \cite{Osterbrock}; $\Sigma(\Omega)$ is the surface density of dust along the line of sight, also output from the \verb"RAMSES" simulations; and $k_{ext}$ is the dust extinction opacity as modelled for the Small Magellanic Cloud (SMC) from \cite{Li_2001} and \cite{Weingartner_2001}. We use these dust extinction opacities because high redshift galaxies, like the SMC,  are expected to have low metallicity. SEDs accounting for absorption are calculated for 100 randomly chosen sightlines to each galaxy at $z=7$ for an observer positioned at infinity, meaning that rays to each star particle for a particular sightline are assumed to be parallel.

Note that while galaxies for which SEDs are calculated without absorption are sampled at multiple time points, specifically up to 100 times following the end of every starburst, galaxies for which SEDs are calculated with absorption are sampled at a single time point ($z=7$). Further, for SEDs with absorption, galaxies are only sampled if they are up to 100 Myr after a star formation peak. Each of these galaxies is sampled 100 times, for 100 randomly chosen lines of sight. Galaxies are only sampled at one time when absorption is included, because while we can reconstruct the star formation histories of these galaxies, we cannot reconstruct their past evolution of the gas.

\subsubsection{Pop III Stars}
\label{sec:popiii}

To model SEDs of Pop III stars, we use the results of \cite{Schaerer2002} and \cite{Zackrisson_2011}. Current predictions for the initial mass functions (IMFs) of Pop III stars are uncertain. While some estimates suggest that Pop III stars are very massive (several hundred $M_\odot$, as in \cite{Bromm2001,2002Abel}), other estimates suggest that certain Pop III stars are much less massive \citep{Tan2004, Stacy2016}. Thus, we employ four different IMFs: three Salpeter IMFs with different upper and lower mass limits ($1<M/M_\odot<100$, $1<M/M_\odot<500$, $50<M/M_\odot<500$), and one log-normal IMF with mass limits $1<M/M_\odot<500$, a characteristic mass $M_c = 10\ M_\odot$, and a dispersion $\sigma = 1\ M_\odot$. We denote these models as IMF A, IMF B, IMF C, and IMF D, respectively. IMFs A and B correspond to Models A and B in \cite{Schaerer2002}, while IMFs C and D correspond to the PopIII.1 and PopIII.2 populations in the Yggdrasil models of \cite{Zackrisson_2011}.

For results without absorption, we use intrinsic, zero-age SEDs for IMFs A and B, and intrinsic SEDs aged 0.01 Myr for IMFs C and D. For results with absorption, we calculate SEDs by replacing the stellar particles in the simulations with zero-age Pop III populations for IMFs A and B, and 0.01 Myr age Pop III populations for IMFs C and D. We then account for absorption as described in \S \ref{sec:single_stripped_stars}. As a result, for SEDs without absorption, we provide a single value for each SED measure for each Pop III IMF, while for SEDs with absorption, we provide a distribution of values (corresponding to the effects of absorption) for each SED measure for each Pop III IMF.  We use only zero-age Pop III stars because most ionizing emission is emitted by the most massive stars, which have very short lifetimes ($\sim 3$ Myr), and the SED characteristics discussed in \S \ref{sec:full_res} do not change significantly for a population of Pop III stars while the most massive stars are still alive. We stress that, since we use cosmological simulations not aimed at simulating Pop III galaxies nor Pop III star formation specifically, the results for Pop III stars with absorption are rather approximate.

\bigskip

\section{Results} \label{sec:full_res}
\subsection{Without Absorption} \label{sec:res}
In this section we examine and compare the intrinsic SEDs of galaxies that do and do not include stripped stars, without any absorption. We also compare these results to intrinsic SEDs for Pop III stellar populations with three different Salpeter IMFs: $1<M/M_\odot<100$ (IMF A), $1<M/M_\odot<500$ (IMF B), $50<M/M_\odot<500$ (IMF C), and one log-normal IMF ($1<M/M_\odot<500$, IMF D). For our comparison we use four metrics.  First, we define a broken power-law, $L_{\lambda} \propto \lambda^{\alpha_i}$, for three wavelength intervals: 240-500 $\text{\AA}$ ($\alpha_1$, hydrogen and neutral helium ionizing), 600-900 $\text{\AA}$ ($\alpha_2$, hydrogen ionizing), and 1200-2000~$\text{\AA}$ ($\alpha_3$). Our $\alpha_3$ is similar to the commonly used UV continuum slope $\beta$ \citep[see e.g.,][]{Shivaei2018, 2016ApJ...833..254S}. We do not analyze the hard HeII-ionizing regime of the SEDs because the ionizing emission rate of HeII-ionizing photons is very sensitive to the treatment of stellar winds, which is uncertain for hot stars (see \citealt{2014ARA&A..52..487S}, \citealt{2020MNRAS.491.4406S}, and in particular \citealt{Gotberg2017} for the effect for stripped stars). 

Second, we define the luminosity ratio, $L_{FUV}/L_{LyC}$, as the ratio of total luminosity in the 1200-2000 $\text{\AA}$ wavelength interval ($L_{FUV}$) to total luminosity in the 240-900 $\text{\AA}$ wavelength interval ($L_{LyC}$). This ratio is similar to the $f_{1500}/f_{900}$ ratio, commonly applied to characterize observed ionizing radiation from LyC-leaking galaxies by comparing observed fluxes at 1500 $\text{\AA}$ and 900 $\text{\AA}$ \citep[see e.g.,][]{2018MNRAS.476L..15V, 2015ApJ...810..107M}.

\begin{figure}
\centering
\includegraphics[width=0.5\textwidth]{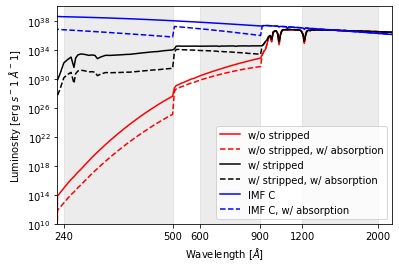}
\cprotect\caption{An example intrinsic SED and SED after absorption (in solid and dashed lines, respectively) for a galaxy with stripped stars (black), a galaxy without stripped stars (red), and a Pop III galaxy with a Salpeter IMF from 50 $M_\odot$ to 500 $M_\odot$ (IMF C, blue). The galaxy has a stellar mass of $4.2 \times 10^5 M_{\odot}$, and is modeled $\sim$ 50 Myr after the most recent starburst.
Shaded grey regions show wavelength intervals used to calculate power-law indices: 240-500 $\text{\AA}$ ($\alpha_1$), 600-900 $\text{\AA}$ ($\alpha_2$), and 1200-2000~$\text{\AA}$ ($\alpha_3$). Pop III SEDs are re-scaled to the same flux level at 1500 $\text{\AA}$ as the SEDs for galaxies without stripped stars, for ease of comparison.}
\label{fig:spectra}
\end{figure}

Figure \ref{fig:spectra} shows an example SED for a galaxy (stellar mass $4.2 \times 10^5 M_{\odot}$, modeled $\sim$ 50 Myr after the most recent starburst) with stripped stars (black), without stripped stars (red), and a Pop III galaxy (IMF C, blue) before (solid) and after (dashed) absorption by intervening gas and dust. We see that stripped stars significantly change spectral shape at ionizing wavelengths, by flattening the spectral slope, but do not affect spectral shape in the FUV regime. We also see that the spectral shape of Pop III galaxies, while bearing more similarity to the spectral shape of galaxies with stripped stars than galaxies without stripped stars, is distinct from both stellar populations. In the sections that follow, we examine this effect in all sampled galaxies.

\subsubsection{Time Evolution of $\alpha_i$}

\begin{figure*}
\centering
\includegraphics[width=\textwidth]{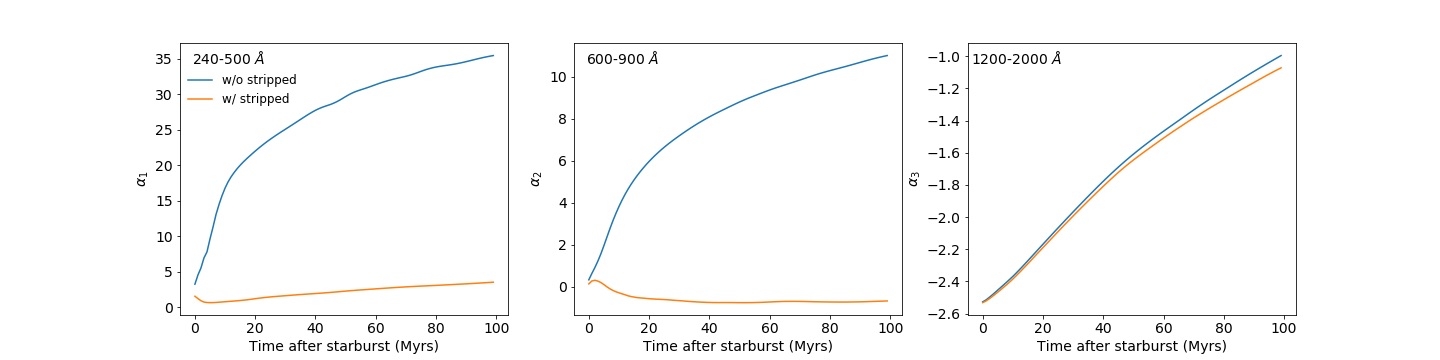}
\caption{An example of the power-law index $\alpha_i$ (defined as the exponent in the power-law: $L_{\lambda} \propto \lambda^{\alpha_i}$) for a single galaxy (stellar mass $8.68 \times 10^4\ M_\odot$) over time, following a recent starburst. The power-law is calculated over three wavelength intervals: 240-500 $\text{\AA}$  ($\alpha_1$), 600-900 $\text{\AA}$ ($\alpha_2$), and 1200-2000~$\text{\AA}$ ($\alpha_3$). Indices for galaxy SEDs that do not include stripped stars are shown in blue while indices for galaxy SEDs that do include stripped stars are shown in orange.}
\label{fig:slopes_ex}
\end{figure*}

Figure \ref{fig:slopes_ex} shows a typical example of the evolution of the power-law index for each wavelength range for a single galaxy as a function of time since the previous starburst. The blue line shows $\alpha_i$ for galaxy SEDs without stripped stars, while the orange line shows $\alpha_i$ for galaxy SEDs with stripped stars. Each panel shows $\alpha_i$ for a different wavelength interval. 

In the 240-500 $\text{\AA}$ and 600-900 $\text{\AA}$ intervals, there is a clear difference in the evolution of power-law index over time when stripped stars are included. For neutral helium-ionizing wavelengths (240-500 $\text{\AA}$), the power-law exponent for the SED that does not include stripped stars rises quickly for approximately the first 10~Myr following the end of a star formation peak, then continues to increase at a slower rate. The initially low values of $\alpha_1$ for these SEDs can be attributed to hard radiation from newly-formed O/B stars, as modelled by \cite{2010ApJS..189..309L}. As these O/B stars reach the end of their lifetime, there is a significant decrease in this hard radiation; $\alpha_1$ increases above 5 after 4 Myr, reaches 20 after 10 Myr and eventually 35 after 100 Myr. While we also see a slight increase in $\alpha_1$ for the SED that does include stripped stars, this rise is constant over time, and the power-law index remains below 5 over 100~Myr.

The behavior of $\alpha_2$ is similar to that of $\alpha_1$ described above. For the galaxy without stripped stars, we again see a sharp increase soon after star formation followed by a less severe ($\alpha_2$ only reaches $\sim 11$), but consistent increase over 100~Myr. In contrast, the power-law for the SED including stripped stars settles at a constant value of $\alpha_2 \approx -0.7$ approximately 20 Myr after the end of star formation. This slight negative value indicates a preference for the lowest wavelength radiation in this interval, which originates from the high temperature of the stripped stars. Overall, Figure \ref{fig:slopes_ex} shows that the harder ionizing emission of stripped stars has a large impact on the SED in the LyC wavelength range, leading to lower values of $\alpha_{1,2}$, especially at later times.

In the 1200-2000 $\text{\AA}$ interval, in line with the similar studies presented in  \S6.2 of \cite{Gotberg2019}, we find that stripped stars do not significantly affect the power-law index of this galaxy's SED. We see a slight difference at later times, where the exponent for the SED with stripped stars is slightly more negative. This difference is small compared to the more prominent difference when stripped stars are included at LyC wavelengths. We expect stripped stars not to have as strong an effect at longer wavelengths, because even though stripped stars have a harder UV slope than the other stars in the stellar population, they are orders of magnitudes fainter in the UV and optical wavelengths. As a result, they are outshone at longer wavelengths, meaning that we do not see their footprint at these wavelengths in full galaxy SEDs \citep{Gotberg2019}.

While above we have calculated power-law indices from luminosity as a function of wavelength, another commonly used definition of the spectral index measures luminosity as a function of frequency. Here, for ease of comparison, we briefly mention corresponding values of $\alpha_i$ over time following a star formation peak, when the power-law is defined as $L_{\nu} \propto \nu^{\alpha_{i,\nu}}$. In the 240-500 $\text{\AA}$ wavelength interval, when stripped stars are not included, we find that the spectral index $\alpha_{1,\nu}$ reaches $\sim -37$ after 100 Myr, while steadily decreasing from $-3.5$ to $-5.5$ when stripped stars are included. In the 600-900 $\text{\AA}$ interval, we find that when stripped stars are included, the power-law settles at a constant value of $\alpha_{2,\nu} \approx -1.3$. These values can be compared to values of $\alpha_{\nu}$ for other ionizing sources, such as AGN, which are expected to be within the range $-2.0 \leq \alpha_{\nu} \leq -1.2$, with some observations indicating slopes of $\alpha_{\nu} \approx -1.7$ \citep[see references in][]{2016MNRAS.456.3354F}. In the 1200-2000 $\text{\AA}$ interval, the power-law evolves from $\sim  0.5$ immediately after the end of star formation to $\sim  -1$ after 100 Myr for both stellar populations.

\subsubsection{Metric Distributions by Luminosity}

\begin{figure}
\centering
\includegraphics[width=0.4\textwidth]{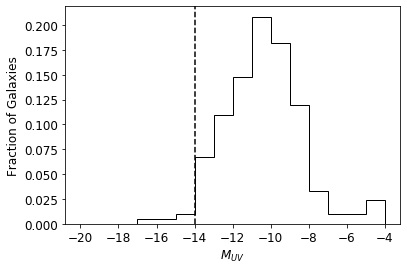}
\caption{Distribution of UV magnitudes for 100 randomly chosen sightlines per galaxy (accounting for absorption by gas and dust) at a redshift $z = 7.$ The vertical line shows UV magnitudes below which galaxies may be detectable by JWST, according to \cite{Behroozi_2020}.}
\label{fig:mag_hist}
\end{figure}

We now turn to the distributions of the four metrics
described at the beginning of this section for simulated galaxies with absolute UV magnitude between $-15$ and $-11$.
The absolute UV magnitude (calculated in the AB magnitude system) was averaged between 1485 and 1515 $\text{\AA}$ to avoid potential fluctuations from spectral features. The complete distribution of UV magnitudes when accounting for gas and dust absorption for 100 randomly chosen sightlines per galaxy at $z = 7$ is shown in Figure \ref{fig:mag_hist}. The dashed vertical line indicates which galaxies we might expect to be observable by JWST, according to \cite{Behroozi_2020}. 

\begin{figure}
\centering
\includegraphics[width=0.45\textwidth]{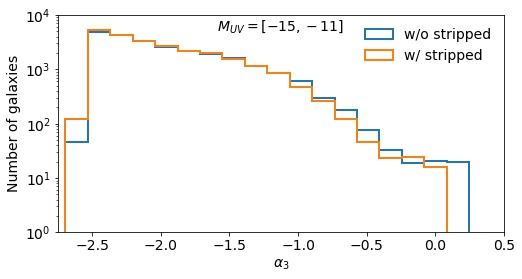}
\caption{Histogram of the power-law index over the 1200-2000 $\text{\AA}$ wavelength interval ($\alpha_3$) for galaxies with UV magnitude between $-15$ and $-11$. Magnitudes were calculated as absolute magnitudes at 1500 $\text{\AA}$. The distributions for galaxy SEDs that include stripped stars are plotted in orange while the distributions for galaxy SEDs that do not include stripped stars are plotted in blue. }
\label{fig:slopes_fuv}
\end{figure}

Figure \ref{fig:slopes_fuv} presents a histogram of power-law indices in the 1200-2000 $\text{\AA}$ wavelength interval for galaxies in the UV magnitude interval mentioned above with (in orange) and without (in blue) stripped stars. These power-law indices (largely concentrated between $-2.5$ and $0.5$) are similar to those cited in studies that measure UV continuum slopes from observations, such as \cite{Shivaei2018}, which give power-law indices clustered between $-2.5$ and $-0.6$ (calculated from 1268-2580 $\text{\AA}$) at $z=2$ (see also \citealt{2016ApJ...831..176B} and \citealt{2019ApJ...881..124M} for measurements of UV continuum slopes at higher redshifts). We find that the distributions of power-law indices of the SEDs for galaxies with and without stripped stars overlap almost completely, suggesting again that stripped stars have little effect on SEDs in these wavelengths.

\begin{figure*}
\centering
\includegraphics[width=\textwidth]{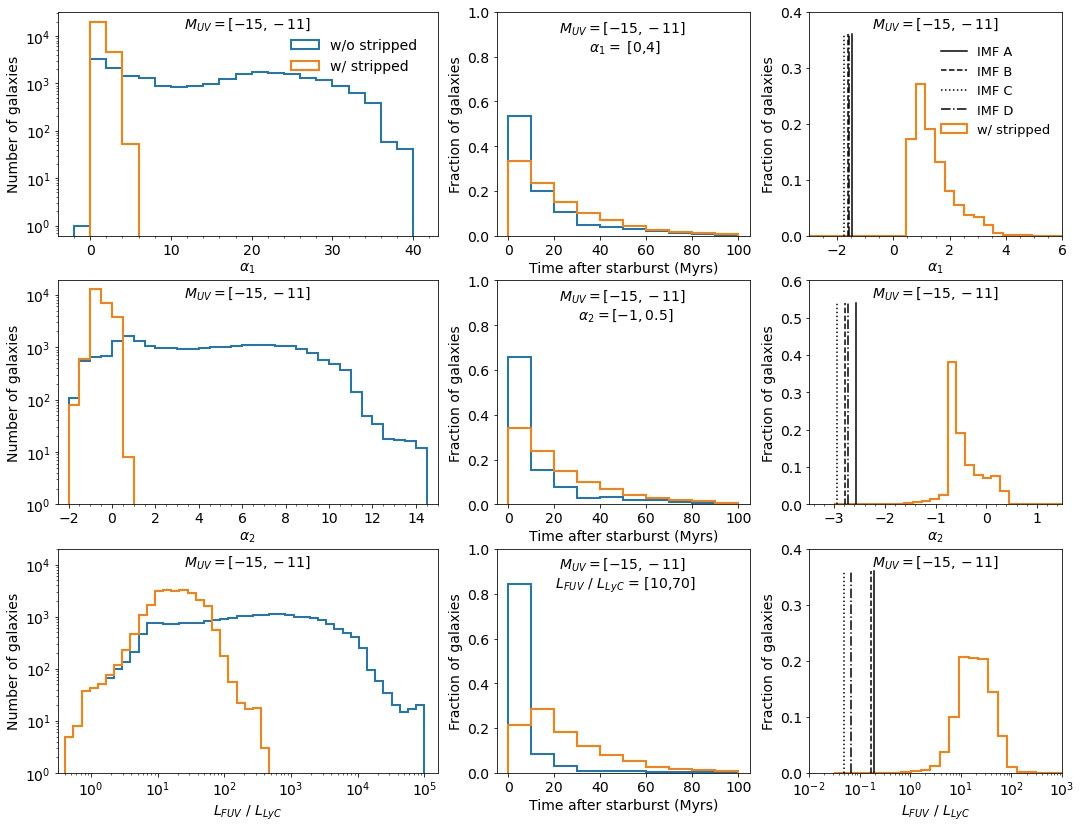}
\caption{Distributions of $\alpha_1$ (top), $\alpha_2$ (middle), and $L_{FUV}/L_{LyC}$ (bottom) for all sampled galaxies with $M_{UV} = [-15,-11]$. \textbf{Left Panel:} Histograms of each metric with galaxies with stripped stars in orange and galaxies without stripped stars in blue. \textbf{Middle Panel:} Distributions of times after starburst for galaxy SEDs with $\alpha_1$ between 0 and 4 (top), $\alpha_2$ between $-1$ and $0.5$ (middle), and $L_{FUV}/L_{LyC}$ between 10 and 70 (bottom), corresponding to the prominent peaks plotted in the left panel. \textbf{Right Panel:} Histograms of each metric for galaxy SEDs that include stripped stars (orange). Vertical lines show values for Pop III populations with IMFs A (solid black line), B (dashed black line), C (dotted black line), and D (dash-dotted black line).
\label{fig:no_abs}}
\end{figure*}

Figure \ref{fig:no_abs} presents results for $\alpha_1$, $\alpha_2$, and $L_{FUV}/L_{LyC}$ for $M_{UV} = [-15,-11]$ in the top, middle, and bottom panels respectively. The left panels are formatted in the same way as Figure \ref{fig:slopes_fuv}. The middle panels of Figure \ref{fig:no_abs} show the distributions of times after starburst for galaxies with SED metrics corresponding to signature peaks from stripped stars in the left panels: $0<\alpha_1<4,-1<\alpha_2<0.5$, and $10 < L_{FUV}/L_{LyC} < 70.$ The right panels present a comparison to Pop III stars for each metric.

We find that $\alpha_1$ and $\alpha_2$ peak far more prominently near zero when stripped stars are included. When stripped stars are not present, we see that more galaxies tend to have larger power-law indices. In the top left panel the distribution of galaxies with stripped stars has a signature peak when $0<\alpha_1<4$. Within this range, galaxies with stripped stars outnumber galaxies without by a factor of $\sim 4.4$. Similarly, we see a signature peak in $\alpha_2$ between $-1$ and $0.5$, where galaxies with stripped star outnumber galaxies without stripped stars by a factor of $8.9$. 

The middle panels of the top two rows in Figure \ref{fig:no_abs} show that the most recent starburst for over 60\% of galaxies with stripped stars that contribute to the peaks in $\alpha_1$ and $\alpha_2$ was more than 10~Myr ago. Galaxies that include stripped stars and are more than 10 Myr past their most recent starburst likely fall within these low $\alpha_1$ and $\alpha_2$ peaks due to the ``delayed'' LyC radiation from stripped stars. For this reason, there is a larger fraction of galaxies older than 10 Myr when stripped stars are included. 

It is somewhat unexpected that a significant proportion of galaxies that do not include stripped stars continue to have low values of $\alpha_1$ and $\alpha_2$ 10 Myr after the most recent starburst, as the lifetime of the O/B stars that produce hard radiation is typically under 10 Myr. This result suggests there may be some ongoing star formation after the main peak has ended.

In the top and center right panels of Figure \ref{fig:no_abs}, we see no overlap in the distribution of $\alpha_1$ and $\alpha_2$ for galaxies with stripped stars and the Pop III models. Interestingly, while all Pop III IMFs produce SEDs with negative values of $\alpha_1$, $\alpha_1$ is positive for all galaxies with stripped stars. As expected, IMF A, with the lowest upper mass limit, has the softest SED, while IMF C, with the highest lower mass limit, has the most negative values for $\alpha_1$ and $\alpha_2$. This trend indicates that we may expect more similarity in the SEDs of galaxies with stripped stars and galaxies with certain low-mass Pop III stars, but that galaxies composed of Pop III stars with masses greater than $100 M_\odot$ have different spectral shapes.

The bottom panels of Figure \ref{fig:no_abs} show results for the ratio of total luminosity between 1200 and 2000 $\text{\AA}$ to total luminosity between 240 and 900 $\text{\AA}$.
 
These results are similar to measurements of $f_{1500}/f_{900}$ flux density ratios from LyC-leaking galaxies. \cite{2018MNRAS.476L..15V}, for example, find a ratio of $\sim 19$ for a star-forming galaxy with stellar mass $1.5 \times 10^9\ M_{\odot}$, while \cite{2015ApJ...810..107M} find a ratio of $\sim 4.0$ for a galaxy with stellar mass $4.8 \times 10^8\ M_{\odot}$. 

The distributions of FUV to LyC luminosity ratios for galaxies with and without stripped stars overlap very closely when $L_{FUV}/L_{LyC} < 3$. These SEDs can be attributed primarily to galaxies less than 10 Myr after star formation peaks, indicating that galaxies with the largest proportion of ionizing radiation relative to FUV radiation are dominated by recent starbursts. The largest peak in the distribution of SEDs including stripped stars occurs for ratios roughly between 10 and 70, and mostly can be attributed to times 10-30 Myr after the most recent peak.

Specifically, we find that $\sim 80\%$ of galaxies that include stripped stars and contribute to ratios between 10 and 70 are more than 10 Myr past the most recent starburst, meaning that we expect these galaxies to be dominated by radiation from stripped stars. 

The difference between the fraction of galaxies with and without stripped stars that fall within the selected range and are over 10~Myr since the last starburst is most dramatic for this metric; $84\%$ of galaxies without stripped stars are less than 10~Myr past their most recent star formation peak. This result suggests even ongoing star formation post-starburst cannot lead to values of $L_{FUV}/L_{LyC}<70$. As a result, SEDs falling within this peak are almost exclusively from galaxies undergoing rapid star formation or containing stripped stars.

In the bottom right panel of Figure \ref{fig:no_abs}, we see that Pop III galaxies, for all four IMFs, have smaller luminosity ratios than what we predict in galaxies with stripped stars. In fact, these ratios are always less than one, suggesting that galaxies dominated by LyC radiation are either starburst galaxies or made up of Pop III stars, rather than dominated by emission from stripped stars, which instead will lead to ratios between 10 and 70.

\begin{figure*}
\centering
\includegraphics[width=\textwidth]{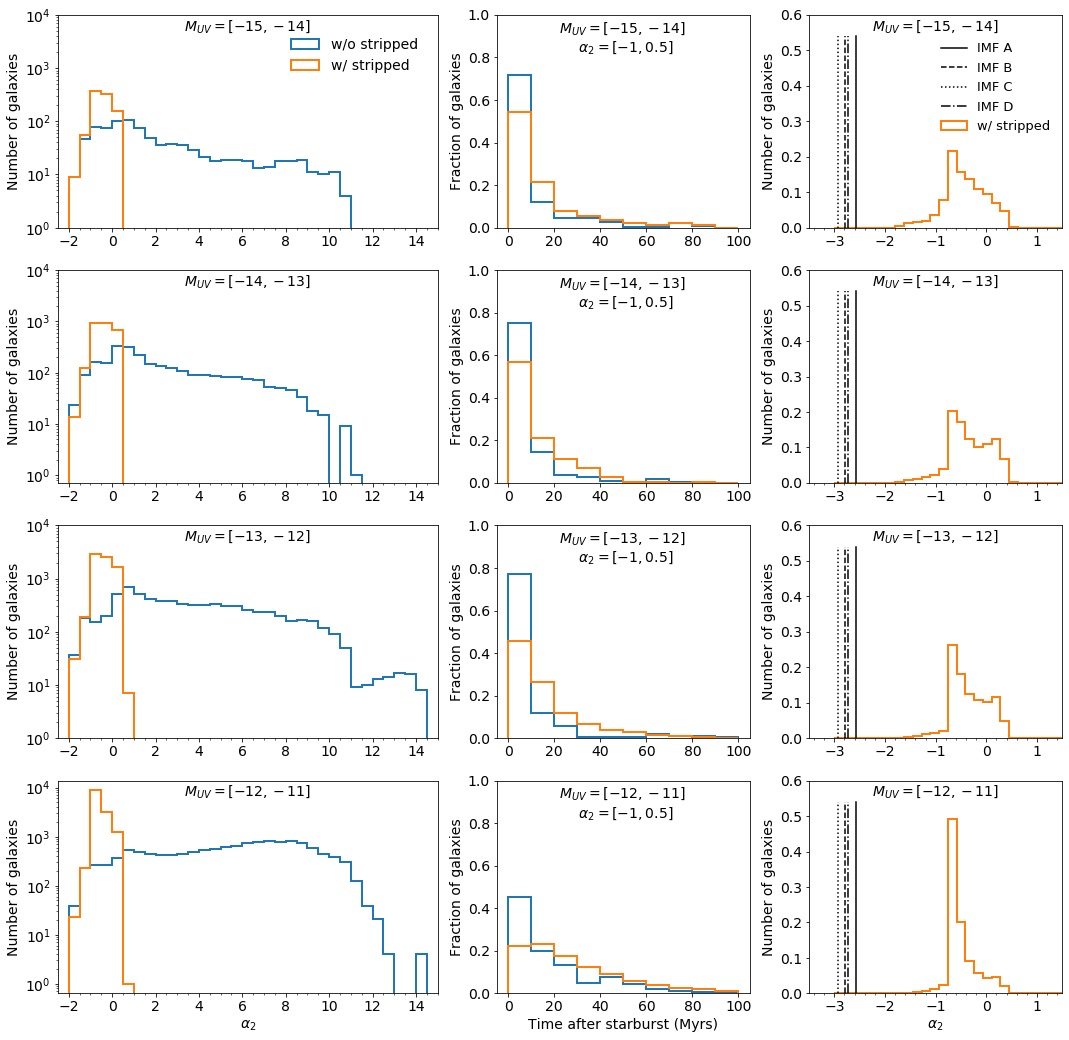}
\caption{\textbf{Left Panel:} Histograms of the power-law index over the 600-900 $\text{\AA}$ wavelength interval ($\alpha_2$) for galaxy SEDs in four magnitude intervals, with the same notation as in the left panels of Figure \ref{fig:slopes_fuv}. \textbf{Middle Panel:} Distributions of times after starburst for galaxy SEDs with power-law index $\alpha_2$ between $-1$ and 0.5, corresponding to the prominent peaks and four magnitude intervals plotted in the left panel. \textbf{Right Panel:} Histograms of power-law index $\alpha_2$ over the 600-900 $\text{\AA}$ wavelength interval in four different magnitude intervals for galaxy SEDs that include stripped stars (orange). Vertical lines show values of $\alpha_2$ for Pop III populations with IMFs A (solid black line), B (dashed black line), C (dotted black line), and D (dash-dotted black line).
\label{fig:slopes_lyc}}
\end{figure*}

We now discuss the dependence of these metrics on UV magnitude, using $\alpha_2$ as a typical example. Figure \ref{fig:slopes_lyc} presents results for power-law indices for the 600-900  $\text{\AA}$ wavelength interval, divided into four magnitude bins: $M_{UV}=[-15,-14], [-14,-13]$, $[-13,-12], [-12,-11]$ mag. Figure \ref{fig:slopes_lyc} is formatted from left to right in the same way as Figure \ref{fig:no_abs}.

For galaxies that do not include stripped stars, we see that the power-law index tends to increase with decreasing luminosity. Higher luminosity galaxies tend to have more recent star formation and therefore have harder ionizing radiation emitted by newly formed O/B stars. The presence of O/B stars leads to a slight peak in galaxy SEDs without stripped stars at smaller values of $\alpha_2$. This peak shifts to the right as luminosity decreases and fewer galaxies have recent star formation. In contrast, we see a prominent peak in the power-laws of galaxies with SEDs that include stripped stars at $\alpha_2$ between $-1$ and 0.5, which remains largely unchanged at all luminosities. Interestingly, when stripped stars are included, we find no galaxies with $\alpha_2 > 1$. For the most luminous magnitude bin, we find that galaxies with stripped stars outnumber galaxies without stripped stars by roughly a factor of 3.5 at these lower values of $\alpha_2$. However, the difference between galaxy SEDs with and without stripped stars is clearest in lower luminosity galaxies, where there are almost two orders of magnitude more galaxies with negative power-law indices when stripped stars are included. In summary, we find that stripped stars strongly effect the distribution of $\alpha_2$ in all considered magnitude bins, and that the effect becomes even more prominent for fainter galaxies. The same effect was observed for distributions of $\alpha_1$ and $L_{FUV}/L_{LyC}$, with both trending towards higher values for lower luminosity galaxies. 

The middle panels of Figure \ref{fig:slopes_lyc} show the fraction of galaxies within the peak in $\alpha_2$ described above ($-1<\alpha_2<0.5$) that are anywhere from 0 to 100~Myr after their last starburst. For the highest luminosity galaxies ($M_{UV} < -14$), where recent star formation is significant, we find that the peaks at smaller $\alpha_2$ are mainly made up of galaxies less than 10 Myr after starburst ($\sim 75 \%$ for SEDs without stripped stars, and $\sim 55 \%$ for SEDs with stripped stars). At lower luminosities, a majority of galaxies that include stripped stars and are within the selected range in $\alpha_2$ are older than 10 Myr. For the least luminous galaxies, for example, 80\% of galaxies with stripped stars that contribute to the shallower $\alpha_2$ are over 10 Myr past their most recent starburst. These results again indicate that we are most likely to see the effects of stripped stars in the lowest luminosity galaxies. However, even for galaxies in the brightest magnitude bin, we are likely to see the effects of stripped stars, as $\sim 45 \%$ of these galaxies are more than 10 Myr after recent star formation.  

The right panels of Figure \ref{fig:slopes_lyc} present a comparison between the distribution of $\alpha_2$ for the SEDs of galaxies with stripped stars to $\alpha_2$ for the SEDs of a population of Pop III stars with IMFs A, B, C, and D, plotted as solid, dashed, dotted, and dash-dotted vertical black lines, respectively.  We see that all Pop III IMFs have harder SEDs than those of galaxies with stripped stars, at all luminosities.

\subsection{Accounting for Absorption} \label{sec:abs}
To more accurately determine $\alpha_i$ and $L_{FUV}/L_{LyC}$ as would be viewed by an observer, we account for absorption by hydrogen, helium, and dust, as a function of frequency. For each galaxy, we choose 100 random lines of sight and position an observer at infinity. Values are calculated only at $z = 7$. Thus, we note that our histograms with absorption present fewer galaxies, particularly at higher luminosities (see Figure \ref{fig:mag_hist}). Specifically, the following results consist of 200 sightlines for $M_{UV} = [-15, - 14]$, 1400 sightlines for $M_{UV} = [-14, - 13]$, 2300 sightlines for $M_{UV} = [-13, - 12]$, and 3200 sightlines for $M_{UV} = [-12, - 11]$. Below we compare our results for galaxies that include emission from only single stellar populations, galaxies that also include emission from stripped stars, and galaxies that include emissions from only zero-age Pop III stars.

\begin{figure*}
\centering
\includegraphics[width=\textwidth]{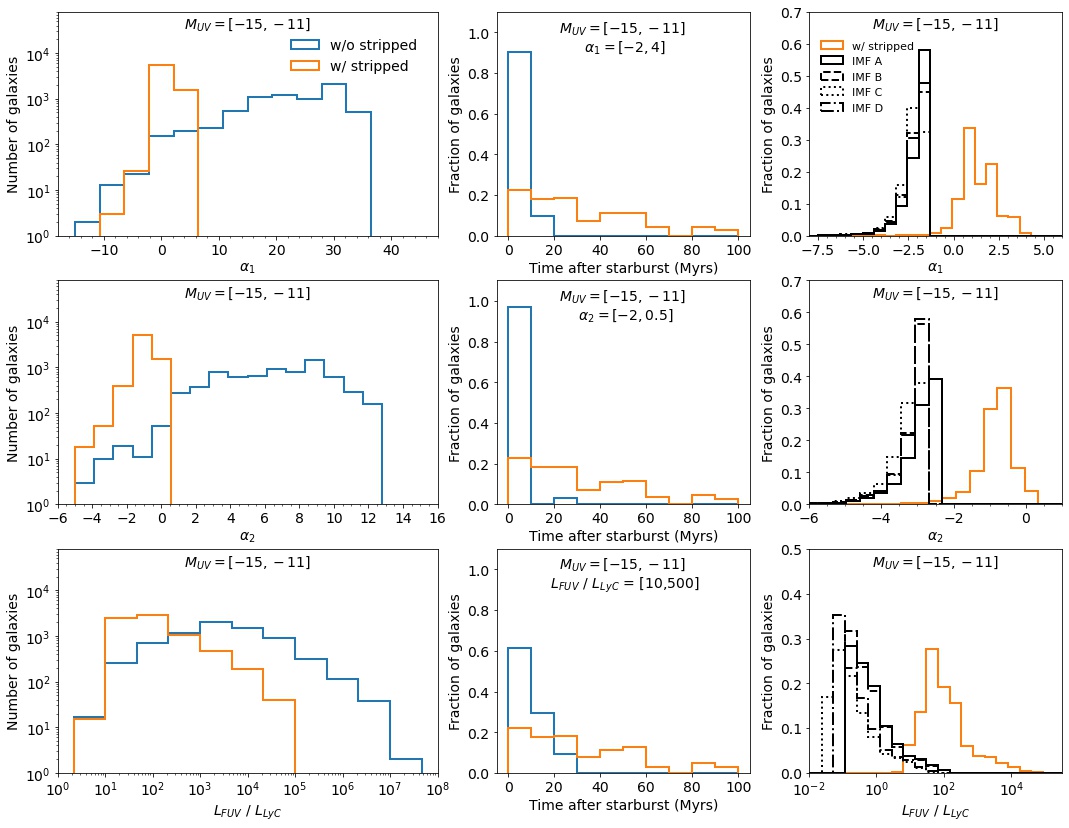}
\caption{Distributions of $\alpha_1$ (top), $\alpha_2$ (middle), and $L_{FUV}/L_{LyC}$ (bottom) for all sampled galaxies with $M_{UV} = [-15,-11]$, calculated for 100 randomly chosen lines of sight per galaxy. \textbf{Left Panel:} Histograms of each metric with galaxies with stripped stars in orange and galaxies without stripped stars in blue. \textbf{Middle Panel:} Distributions of times after starburst for galaxy SEDs with $\alpha_1$ between $-2$ and 4 (top), $\alpha_2$ between $-2$ and $0.5$ (middle) , and $L_{FUV}/L_{LyC}$ between 10 and 500 (bottom), corresponding to the prominent peaks plotted in the left panel. \textbf{Right Panel:} Histograms of each metric for galaxy SEDs that include stripped stars (orange) and Pop III populations with IMFs A (solid black line), B (dashed black line), C (dotted black line), and D (dash-dotted black line)}.
\label{fig:abs}
\end{figure*}

Figure \ref{fig:abs} is the same as Figure \ref{fig:no_abs}, except it presents results accounting for absorption by intervening gas and dust (see \S \ref{sec:single_stripped_stars}, \S \ref{sec:popiii}). For all metrics, we again see a consistent peak at shallower power-law indices for galaxy SEDs that do include stripped stars. We see a signature peak in galaxies with stripped stars for $-2 < \alpha_1 < 4$, $-2 < \alpha_2 < 0.5$, and $10 < L_{FUV}/L_{LyC} < 500$. Compared to results that did not account for absorption, we also find that absorption tends to cause a slight trend towards negative values of $\alpha_1$ and $\alpha_2$, due to the tendency of H and He to absorb more at lower wavelengths over these intervals. For $L_{FUV}/L_{LyC}$, we see a shift towards larger ratios when we account for absorption. This shift is a result of H and He absorption at ionizing wavelengths.

With absorption we find that $\sim80\%$ of galaxies that include stripped stars and make up the prominent peaks for each of the metrics are over 10 Myr after starburst. This proportion is larger than without absorption, because within the first 10~Myr after a starburst, before feedback has time to remove gas from the birth cloud, most LyC radiation will be absorbed. On the other hand, ionizing photons from stripped stars, which begin to form at an approximately 10 Myr delay following the end of star formation, will be able to escape and produce the hard SEDs that make up the peaks in the distributions in the left panels of Figure \ref{fig:abs}. In contrast, the same ranges of power-law indices for SEDs without stripped stars are composed almost exclusively of galaxies experiencing recent star formation peaks. This result suggests that absorption of LyC photons in the still gas-rich regions undergoing star formation after a galaxy's star formation peak prevents these regions from significantly hardening their galaxy's SED as they did in \S \ref{sec:res}.

The right panels of Figure \ref{fig:abs} show a comparison between SEDs from galaxies with stripped stars and galaxies composed of Pop III populations of various IMFs, where we have accounted for absorption. Similar to results from galaxies without absorption, we find that Pop III stars produce the hardest ionizing spectra. For Pop III populations with an IMF from 1-100 $M_\odot$ (IMF A), we see a very slight overlap with the stripped-star distribution of $\alpha_1$ and $\alpha_2$. The luminosity ratio distributions show the most overlap between Pop III galaxies and stripped-star galaxies. Still, the Pop III and stripped-star populations are clearly distinct.

\begin{figure*}
\centering
\includegraphics[width=0.7\textwidth]{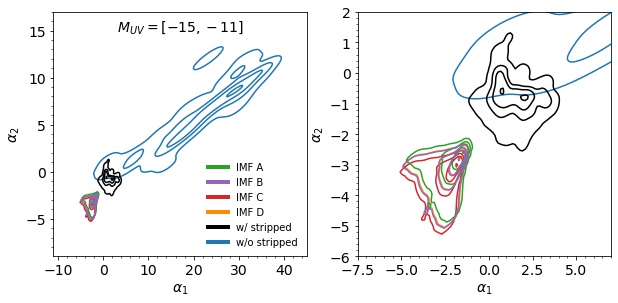}
\caption{\textbf{Left Panel:} Contour plot of the number of galaxies in the $\alpha_1 - \alpha_2$ plane for galaxies with stripped stars (black), galaxies without stripped stars (blue), Pop III galaxies with IMF A (green), IMF B (purple), IMF C (red), and IMF D (orange). Contour lines contain 5\%, 50\%, 75\%, and 95\% of cases from inside out. \textbf{Right Panel:} Same as the left panel but zoomed in on galaxies with stripped stars and galaxies composed of Pop III stars.
\label{fig:alpha_density}}
\end{figure*}

Figure \ref{fig:alpha_density} presents contour plots for each set of stellar populations for results with absorption in the $\alpha_1 - \alpha_2$ plane, with contour lines containing 5\%, 50\%, 75\%, and 95\% of galaxies from inside out. The left panel displays galaxies with stripped stars, galaxies without stripped stars, and Pop III galaxies with IMFs A, B, C, and D, colored black, blue, green, purple, red, and orange, respectively. The right panel is zoomed in on galaxies composed of Pop III stars and galaxies with stripped stars. We see a clear separation between the high-density regions of galaxies with and without stripped stars.

Specifically, we see no overlap between 95\% of galaxies with stripped stars and 75\% of galaxies without stripped stars. We also find that Pop III galaxies are distinct from galaxies with stripped stars. For all four IMFs, we find that Pop III galaxies have more negative values of $\alpha_1$ and $\alpha_2$ than galaxies with stripped stars. Contour lines for IMFs B and D, which have the same upper and lower mass limits ($1<M/M_\odot<500$), but different mass distributions (IMF B uses a Salpeter slope while IMF D is log-normal with $M_c = 10\ M_\odot$ and $\sigma = 1\ M_\odot$), overlap almost completely. This overlap suggests that the maximum mass plays the most significant role in determining the power-law indices for Pop III stellar populations. We also note that a small fraction of Pop III galaxies ($0.2\%$), galaxies with stripped stars ($0.1\%$), and galaxies without stripped stars ($0.02\%$) have $\alpha_2 > \alpha_1$. This effect is caused by strong absorption of radiation in the $\alpha_1$ interval over particular lines of sight, resulting in power-law indices that appear to characterize particularly hard spectra.

\section{Summary and Conclusions} \label{sec:conclusion}
Our findings can be summarized as follows:

\begin{enumerate}
    \item Stripped stars have little effect on SEDs in the 1200-2000  $\text{\AA}$ interval (Figure \ref{fig:slopes_fuv}). This result is expected, given that stripped stars emit most at ionizing wavelengths.
    \item In the 600-900 $\text{\AA}$ interval, galaxy SEDs with stripped stars have signature power-law indices concentrated between $-2$ and 0.5, with no galaxies exhibiting values of $\alpha_2 > 1$ (Figures \ref{fig:no_abs}, \ref{fig:abs}).  We see a clear difference between SEDs with and without stripped stars. 
    When we account for absorption we find that $\sim 80\% $ of galaxies contributing to power-law indices between $-2$ and 0.5 are more than 10 Myr from the end of their most recent starburst. The length of time since the last starburst for these galaxies suggests that their SEDs are dominated by ionizing emission from stripped stars rather than O/B stars formed in recent starburts. 
    Our results suggest that if the SED of a galaxy has a power-law index between 600 and 900 $\text{\AA}$ greater than one, that galaxy would not contain any stripped stars.
    \item In the 240-500 $\text{\AA}$ interval, galaxy SEDs with stripped stars have signature power-law indices concentrated between $-2$ and 4 (Figures \ref{fig:no_abs}, \ref{fig:abs}), with no galaxies exhibiting values of $\alpha_1 > 5$. 
    As in the previous point, when we account for absorption a majority of galaxies with power-law indices between $-2$ and 4 are more than 10 Myr after the most recent starburst. This result again indicates that ionizing radiation from these galaxies is likely to be dominated by stripped stars. 
    \item In the $\alpha_1 - \alpha_2$ plane, as in Figure \ref{fig:alpha_density}, we find a clear distinction between galaxies with and without stripped stars. Specifically, we find that there is no overlap between 95\% of galaxies with stripped stars and 75\% of galaxies without stripped stars. 
    \item 
    $L_{FUV}/L_{LyC}$ peaks between 10 and 500 for galaxy SEDs that include stripped stars, while the distribution of values for galaxy SEDs without stripped stars sits somewhat higher. 
    Approximately 80\% of galaxies that contain stripped stars and contribute to luminosity ratios between 10 and 500 are over 10 Myr past their most recent starburst (Figures \ref{fig:no_abs}, \ref{fig:abs}). As above, this result suggests that the SEDs of these galaxies will be dominated by stripped stars.
    \item There is little similarity in SEDs between galaxies with stripped stars and Pop III populations with IMFs A, B, C, or D. This is clearest when $\alpha_1$ and $\alpha_2$ are taken together, as in Figure \ref{fig:alpha_density}. These results indicate that it is likely possible to distinguish between galaxies containing stripped stars and Pop III stars in most cases. More robust comparisons between Pop III stars and stripped stars will require a better understanding of the Pop III IMF and more detailed models of early galaxies containing Pop III stars.
    \item While differences between each stellar population are clear at all luminosities, we see the greatest differences between galaxies with and without stripped stars at lower luminosities (Figure \ref{fig:slopes_lyc}). Specifically, we see a trend towards higher values of $\alpha_1$,  $\alpha_2$, and $L_{FUV}/L_{LyC}$ for galaxies without stripped stars as luminosity decreases, while metric distributions remain largely unchanged as a function of luminosity when stripped stars are present. We also find that an increasing fraction of galaxies corresponding to signature peaks from stripped stars in each metric are more than 10 Myrs after starburst as luminosity decreases, reaching $\sim 90\%$ for the lowest luminosity galaxies.
\end{enumerate}

\cite{Behroozi_2020} found that galaxies at a UV magnitude less than $-17$ (in the AB system) were likely to be detected by JWST with high confidence ($> 85\%$) at redshifts up to 13.5, and indicated that galaxies at a UV magnitude less than $-14$ may be detectable at lower confidence. Figure \ref{fig:slopes_lyc} suggests that galaxies one to two magnitudes dimmer would have the most distinctive signatures of stripped stars being present. However, Figure \ref{fig:slopes_lyc} still suggests that smaller power-law indices and smaller luminosity ratios are likely to indicate that stripped stars are present even for $M_{UV}$ between $-15$ and $-14$.

From $z = 7$ to $z = 6$, we expect little change in the galaxy luminosity function, particularly at low luminosities \citep{Trac_2015}. At $z = 6$ galaxies with $M_{UV}$ between $-15$ and $-14$ may be detectable by certain JWST ultra-deep surveys or if gravitationally lensed \citep{2013JCAP...12..017M}. 

We also expect the ionizing spectrum to leave imprints on the emission line properties of galaxies. For example, binary populations present a possible explanation for observed high emission line ratios [O III]/H$\beta$ in distant, low-mass, star-forming galaxies \citep{2014MNRAS.444.3466S}, as products of binary evolution tend increase emission line strengths for older stellar populations, relative to those predicted from single-stars. \cite{2018MNRAS.477..904X} modeled and characterized nebular line ratios between single stellar populations and binary populations, and found that binary populations enhance emission line strength, particularly on the metal lines [N II], [S II], [O III] and [O I], for stellar populations older than 10 Myrs. Future work can use our comparison of the differences in the SEDs of galaxies made up of different stellar populations to predict the impact of these differences on emission line properties and other observational signatures.

With the launch of JWST, we will be able to observe faint, low-mass galaxies during cosmic reionization. We predict that stripped stars dramatically increase the ionizing emission from the bulk of these galaxies, which, for example, could significantly affect the morphology of the nebular spectrum. 
We have also shown that stripped stars and Pop III stars appear to have completely different spectral hardness in the ionizing regime. Therefore, it may be possible to use the hardness of ionizing radiation as a key for revealing what sources in a stellar population are responsible for the emitted ionizing radiation. Our results highlight the importance of better understanding the spectral hardness of various different ionizing sources and their effect on observable quantities.

\acknowledgments
We thank Taysun Kimm for kindly sharing
the simulation data.
Computing resources were in part provided by the NASA High-
End Computing (HEC) Program through the NASA Advanced
Supercomputing (NAS) Division at Ames Research Center.
A.S. is supported by the NSF Graduate Research Fellowship Program under Grant No. DGE-1656466.
The research is supported in part by NASA grant 80NSSC18K1101. Y.G. acknowledges support from NASA through the NASA Hubble Fellowship Program grant \#HST-HF2-51457.001-A awarded by the Space Telescope Science Institute, which is operated by the Association of Universities for Research in Astronomy, Inc., for NASA, under contract NAS5-26555.

\vspace{5mm}

\clearpage

\bibliography{bib}

\begin{thebibliography}{}
\expandafter\ifx\csname natexlab\endcsname\relax\def\natexlab#1{#1}\fi
\providecommand{\url}[1]{\href{#1}{#1}}

\bibitem[{{Abdul-Masih} {et~al.}(2019){Abdul-Masih}, {Sana}, {Sundqvist},
  {Mahy}, {Menon}, {Almeida}, {De Koter}, {de Mink}, {Justham}, {Langer},
  {Puls}, {Shenar}, \& {Tramper}}]{2019ApJ...880..115A}
{Abdul-Masih}, M., {Sana}, H., {Sundqvist}, J., {et~al.} 2019, \apj, 880, 115

\bibitem[{{Abel} {et~al.}(2002){Abel}, {Bryan}, \& {Norman}}]{2002Abel}
{Abel}, T., {Bryan}, G.~L., \& {Norman}, M.~L. 2002, Science, 295, 93

\bibitem[{{Almeida} {et~al.}(2017){Almeida}, {Sana, H.}, {Taylor, W.},
  {Barb\'a, R.}, {Bonanos, A. Z.}, {Crowther, P.}, {Damineli, A.}, {de Koter,
  A.}, {de Mink, S. E.}, {Evans, C. J.}, {Gieles, M.}, {Grin, N. J.},
  {H\'enault-Brunet, V.}, {Langer, N.}, {Lennon, D.}, {Lockwood, S.},
  {Ma\'{\i}z Apell\'aniz, J.}, {Moffat, A. F. J.}, {Neijssel, C.}, {Norman,
  C.}, {Ram\'{\i}rez-Agudelo, O. H.}, {Richardson, N. D.}, {Schootemeijer, A.},
  {Shenar, T.}, {Soszy\'{}nski, I.}, {Tramper, F.}, \& {Vink, J.
  S.}}]{Almeida2017}
{Almeida}, {Sana, H.}, {Taylor, W.}, {et~al.} 2017, A\&A, 598, A84.
\newblock \url{https://doi.org/10.1051/0004-6361/201629844}

\bibitem[{Behroozi {et~al.}(2020)Behroozi, Conroy, Wechsler, Hearin, Williams,
  Moster, Yung, Somerville, Gottlöber, Yepes, \& Endsley}]{Behroozi_2020}
Behroozi, P., Conroy, C., Wechsler, R.~H., {et~al.} 2020, Monthly Notices of
  the Royal Astronomical Society, 499, 5702.
\newblock \url{https://doi.org/10.1093/mnras/staa3164}

\bibitem[{{Berg} {et~al.}(2019){Berg}, {Chisholm}, {Erb}, {Pogge}, {Henry}, \&
  {Olivier}}]{2019ApJ...878L...3B}
{Berg}, D.~A., {Chisholm}, J., {Erb}, D.~K., {et~al.} 2019, \apjl, 878, L3

\bibitem[{Bertelli {et~al.}(1993)Bertelli, Bressan, Chiosi, \&
  Fagotto}]{Bertelli_1993}
Bertelli, G., Bressan, A., Chiosi, C., \& Fagotto, F. 1993, in New Aspects of
  Magellanic Cloud Research, ed. B.~Baschek, G.~Klare, \& J.~Lequeux (Berlin,
  Heidelberg: Springer Berlin Heidelberg), 362--363

\bibitem[{{Bertelli} {et~al.}(1994){Bertelli}, {Bressan}, {Chiosi}, {Fagotto},
  \& {Nasi}}]{Bertelli_1994}
{Bertelli}, G., {Bressan}, A., {Chiosi}, C., {Fagotto}, F., \& {Nasi}, E. 1994,
  \aaps, 106, 275

\bibitem[{{Bouwens} {et~al.}(2016){Bouwens}, {Smit}, {Labb{\'e}}, {Franx},
  {Caruana}, {Oesch}, {Stefanon}, \& {Rasappu}}]{2016ApJ...831..176B}
{Bouwens}, R.~J., {Smit}, R., {Labb{\'e}}, I., {et~al.} 2016, \apj, 831, 176

\bibitem[{{Bromm} {et~al.}(2001){Bromm}, {Kudritzki}, \& {Loeb}}]{Bromm2001}
{Bromm}, V., {Kudritzki}, R.~P., \& {Loeb}, A. 2001, \apj, 552, 464

\bibitem[{{Byler} {et~al.}(2019){Byler}, {Dalcanton}, {Conroy}, {Johnson},
  {Choi}, {Dotter}, \& {Rosenfield}}]{2019AJ....158....2B}
{Byler}, N., {Dalcanton}, J.~J., {Conroy}, C., {et~al.} 2019, \aj, 158, 2

\bibitem[{{Carr} {et~al.}(1984){Carr}, {Bond}, \&
  {Arnett}}]{1984ApJ...277..445C}
{Carr}, B.~J., {Bond}, J.~R., \& {Arnett}, W.~D. 1984, \apj, 277, 445

\bibitem[{{Chen} {et~al.}(2015){Chen}, {Woods}, {Yungelson}, {Gilfanov}, \&
  {Han}}]{2015MNRAS.453.3024C}
{Chen}, H.-L., {Woods}, T.~E., {Yungelson}, L.~R., {Gilfanov}, M., \& {Han}, Z.
  2015, \mnras, 453, 3024

\bibitem[{{Crowther} {et~al.}(2016){Crowther}, {Caballero-Nieves}, {Bostroem},
  {Ma{\'\i}z Apell{\'a}niz}, {Schneider}, {Walborn}, {Angus}, {Brott},
  {Bonanos}, {de Koter}, {de Mink}, {Evans}, {Gr{\"a}fener}, {Herrero},
  {Howarth}, {Langer}, {Lennon}, {Puls}, {Sana}, \&
  {Vink}}]{2016MNRAS.458..624C}
{Crowther}, P.~A., {Caballero-Nieves}, S.~M., {Bostroem}, K.~A., {et~al.} 2016,
  \mnras, 458, 624

\bibitem[{{de Mink} {et~al.}(2009){de Mink}, {Cantiello}, {Langer}, {Pols},
  {Brott}, \& {Yoon}}]{2009A&A...497..243D}
{de Mink}, S.~E., {Cantiello}, M., {Langer}, N., {et~al.} 2009, \aap, 497, 243

\bibitem[{{Eldridge} {et~al.}(2017){Eldridge}, {Stanway}, {Xiao}, {McClelland
  }, {Taylor}, {Ng}, {Greis}, \& {Bray}}]{Eldridge2017}
{Eldridge}, J.~J., {Stanway}, E.~R., {Xiao}, L., {et~al.} 2017, \pasa, 34, e058

\bibitem[{{Feltre} {et~al.}(2016){Feltre}, {Charlot}, \&
  {Gutkin}}]{2016MNRAS.456.3354F}
{Feltre}, A., {Charlot}, S., \& {Gutkin}, J. 2016, \mnras, 456, 3354

\bibitem[{G{\"o}tberg {et~al.}(2019)G{\"o}tberg, de~Mink, Groh, Leitherer, \&
  Norman}]{Gotberg2019}
G{\"o}tberg, Y., de~Mink, S.~E., Groh, J.~H., Leitherer, C., \& Norman, C.
  2019, \aap, 629, A134

\bibitem[{{Groh} {et~al.}(2008){Groh}, {Oliveira}, \& {Steiner}}]{Groh2008}
{Groh}, J.~H., {Oliveira}, A.~S., \& {Steiner}, J.~E. 2008, \aap, 485, 245

\bibitem[{Götberg {et~al.}(2017)Götberg, de~Mink, \& Groh}]{Gotberg2017}
Götberg, Y., de~Mink, S.~E., \& Groh, J.~H. 2017, Astronomy \& Astrophysics,
  608, A11.
\newblock \url{http://dx.doi.org/10.1051/0004-6361/201730472}

\bibitem[{Hahn \& Abel(2011)}]{Hahn_Abel_2011}
Hahn, O., \& Abel, T. 2011, Monthly Notices of the Royal Astronomical Society,
  415, 2101.
\newblock \url{https://doi.org/10.1111/j.1365-2966.2011.18820.x}

\bibitem[{{Hainich} {et~al.}(2015){Hainich}, {Pasemann}, {Todt}, {Shenar},
  {Sander}, \& {Hamann}}]{2015A&A...581A..21H}
{Hainich}, R., {Pasemann}, D., {Todt}, H., {et~al.} 2015, \aap, 581, A21

\bibitem[{{Hillier} \& {Miller}(1998)}]{Hellier_1998}
{Hillier}, D.~J., \& {Miller}, D.~L. 1998, \apj, 496, 407

\bibitem[{Kimm \& Cen(2014)}]{KimmCen_2014}
Kimm, T., \& Cen, R. 2014, The Astrophysical Journal, 788, 121

\bibitem[{{Kippenhahn} \& {Weigert}(1967)}]{Kippenhahn67}
{Kippenhahn}, R., \& {Weigert}, A. 1967, \zap, 65, 251

\bibitem[{{Kobulnicky} \& {Fryer}(2007)}]{Kobulnicky2007}
{Kobulnicky}, H.~A., \& {Fryer}, C.~L. 2007, \apj, 670, 747

\bibitem[{Komatsu {et~al.}(2011)Komatsu, Smith, Dunkley, Bennett, Gold,
  Hinshaw, Jarosik, Larson, Nolta, Page, Spergel, Halpern, Hill, Kogut, Limon,
  Meyer, Odegard, Tucker, Weiland, Wollack, \& Wright}]{Komatsu_2011}
Komatsu, E., Smith, K.~M., Dunkley, J., {et~al.} 2011, The Astrophysical
  Journal Supplement Series, 192, 18

\bibitem[{Kroupa(2001)}]{Kroupa2001}
Kroupa, P. 2001, Monthly Notices of the Royal Astronomical Society, 322, 231.
\newblock \url{https://doi.org/10.1046/j.1365-8711.2001.04022.x}

\bibitem[{{Kub{\'a}tov{\'a}} {et~al.}(2019){Kub{\'a}tov{\'a}}, {Sz{\'e}csi},
  {Sander}, {Kub{\'a}t}, {Tramper}, {Krti{\v{c}}ka}, {Kehrig}, {Hamann},
  {Hainich}, \& {Shenar}}]{2019A&A...623A...8K}
{Kub{\'a}tov{\'a}}, B., {Sz{\'e}csi}, D., {Sander}, A.~A.~C., {et~al.} 2019,
  \aap, 623, A8

\bibitem[{{Leitherer} {et~al.}(2010){Leitherer}, {Ortiz Ot{\'a}lvaro},
  {Bresolin}, {Kudritzki}, {Lo Faro}, {Pauldrach}, {Pettini}, \&
  {Rix}}]{2010ApJS..189..309L}
{Leitherer}, C., {Ortiz Ot{\'a}lvaro}, P.~A., {Bresolin}, F., {et~al.} 2010,
  \apjs, 189, 309

\bibitem[{{Leitherer} {et~al.}(1999){Leitherer}, {Schaerer}, {Goldader},
  {Delgado}, {Robert}, {Kune}, {de Mello}, {Devost}, \&
  {Heckman}}]{Starburst99}
{Leitherer}, C., {Schaerer}, D., {Goldader}, J.~D., {et~al.} 1999, \apjs, 123,
  3

\bibitem[{{Levesque} {et~al.}(2012){Levesque}, {Leitherer}, {Ekstrom},
  {Meynet}, \& {Schaerer}}]{2012ApJ...751...67L}
{Levesque}, E.~M., {Leitherer}, C., {Ekstrom}, S., {Meynet}, G., \& {Schaerer},
  D. 2012, \apj, 751, 67

\bibitem[{Li \& Draine(2001)}]{Li_2001}
Li, A., \& Draine, B.~T. 2001, The Astrophysical Journal, 554, 778

\bibitem[{Ma {et~al.}(2016)Ma, Hopkins, Kasen, Quataert, Faucher-Giguère,
  Kereš, Murray, \& Strom}]{Ma_2016}
Ma, X., Hopkins, P.~F., Kasen, D., {et~al.} 2016, Monthly Notices of the Royal
  Astronomical Society, 459, 3614.
\newblock \url{https://doi.org/10.1093/mnras/stw941}

\bibitem[{{Maeder}(1987)}]{1987A&A...178..159M}
{Maeder}, A. 1987, \aap, 178, 159

\bibitem[{{Marigo} {et~al.}(2008){Marigo}, {Girardi}, {Bressan}, {Groenewegen},
  {Silva}, \& {Granato}}]{Marigo_2008}
{Marigo}, P., {Girardi}, L., {Bressan}, A., {et~al.} 2008, \aap, 482, 883

\bibitem[{{Martins} {et~al.}(2009){Martins}, {Hillier}, {Bouret}, {Depagne},
  {Foellmi}, {Marchenko}, \& {Moffat}}]{2009A&A...495..257M}
{Martins}, F., {Hillier}, D.~J., {Bouret}, J.~C., {et~al.} 2009, \aap, 495, 257

\bibitem[{{Mashian} \& {Loeb}(2013)}]{2013JCAP...12..017M}
{Mashian}, N., \& {Loeb}, A. 2013, \jcap, 2013, 017

\bibitem[{Mason {et~al.}(2009)Mason, Hartkopf, Gies, Henry, \&
  Helsel}]{Mason_2009}
Mason, B.~D., Hartkopf, W.~I., Gies, D.~R., Henry, T.~J., \& Helsel, J.~W.
  2009, The Astronomical Journal, 137, 3358.
\newblock \url{https://doi.org/10.1088\%2F0004-6256\%2F137\%2F2\%2F3358}

\bibitem[{{Matthee} {et~al.}(2019){Matthee}, {Sobral}, {Boogaard},
  {R{\"o}ttgering}, {Vallini}, {Ferrara}, {Paulino-Afonso}, {Boone},
  {Schaerer}, \& {Mobasher}}]{2019ApJ...881..124M}
{Matthee}, J., {Sobral}, D., {Boogaard}, L.~A., {et~al.} 2019, \apj, 881, 124

\bibitem[{Moe \& Di~Stefano(2017)}]{Moe_2017}
Moe, M., \& Di~Stefano, R. 2017, The Astrophysical Journal Supplement Series,
  230, 15.
\newblock \url{http://dx.doi.org/10.3847/1538-4365/aa6fb6}

\bibitem[{{Mostardi} {et~al.}(2015){Mostardi}, {Shapley}, {Steidel}, {Trainor},
  {Reddy}, \& {Siana}}]{2015ApJ...810..107M}
{Mostardi}, R.~E., {Shapley}, A.~E., {Steidel}, C.~C., {et~al.} 2015, \apj,
  810, 107

\bibitem[{{Nanayakkara} {et~al.}(2019){Nanayakkara}, {Brinchmann}, {Boogaard},
  {Bouwens}, {Cantalupo}, {Feltre}, {Kollatschny}, {Marino}, {Maseda},
  {Matthee}, {Paalvast}, {Richard}, \& {Verhamme}}]{2019A&A...624A..89N}
{Nanayakkara}, T., {Brinchmann}, J., {Boogaard}, L., {et~al.} 2019, \aap, 624,
  A89

\bibitem[{{{\"O}pik}(1924)}]{Opik1924}
{{\"O}pik}, E. 1924, Astronomische Nachrichten, 221, 223

\bibitem[{{Osterbrock} \& {Ferland}(2006)}]{Osterbrock}
{Osterbrock}, D.~E., \& {Ferland}, G.~J. 2006, {Astrophysics of gaseous nebulae
  and active galactic nuclei} (University Science Books)

\bibitem[{{Pauldrach} {et~al.}(2001){Pauldrach}, {Hoffmann}, \&
  {Lennon}}]{Pauldrach_2001}
{Pauldrach}, A.~W.~A., {Hoffmann}, T.~L., \& {Lennon}, M. 2001, \aap, 375, 161

\bibitem[{{Paxton} {et~al.}(2011){Paxton}, {Bildsten}, {Dotter}, {Herwig},
  {Lesaffre}, \& {Timmes}}]{Paxton2011}
{Paxton}, B., {Bildsten}, L., {Dotter}, A., {et~al.} 2011, \apjs, 192, 3

\bibitem[{{Paxton} {et~al.}(2013){Paxton}, {Cantiello}, {Arras}, {Bildsten},
  {Brown}, {Dotter}, {Mankovich}, {Montgomery}, {Stello}, {Timmes}, \&
  {Townsend}}]{Paxton2013}
{Paxton}, B., {Cantiello}, M., {Arras}, P., {et~al.} 2013, \apjs, 208, 4

\bibitem[{Paxton {et~al.}(2015)Paxton, Marchant, Schwab, Bauer, Bildsten,
  Cantiello, Dessart, Farmer, Hu, Langer, Townsend, Townsley, \&
  Timmes}]{Paxton2015}
Paxton, B., Marchant, P., Schwab, J., {et~al.} 2015, The Astrophysical Journal
  Supplement Series, 220, 15.
\newblock \url{https://doi.org/10.1088\%2F0067-0049\%2F220\%2F1\%2F15}

\bibitem[{{Pols}(1994)}]{Pols1994}
{Pols}, O.~R. 1994, \aap, 290, 119

\bibitem[{{Rees}(1978)}]{1978Natur.275...35R}
{Rees}, M.~J. 1978, \nat, 275, 35

\bibitem[{Rosdahl {et~al.}(2018)Rosdahl, Katz, Blaizot, Kimm, Michel-Dansac,
  Garel, Haehnelt, Ocvirk, \& Teyssier}]{Rosdahl_2018}
Rosdahl, J., Katz, H., Blaizot, J., {et~al.} 2018, Monthly Notices of the Royal
  Astronomical Society, 479, 994.
\newblock \url{https://doi.org/10.1093/mnras/sty1655}

\bibitem[{{Rosen} \& {Bregman}(1995)}]{Rosen1995}
{Rosen}, A., \& {Bregman}, J.~N. 1995, \apj, 440, 634

\bibitem[{Rydberg {et~al.}(2013)Rydberg, Zackrisson, Lundqvist, \&
  Scott}]{Rydberg_2013}
Rydberg, C.-E., Zackrisson, E., Lundqvist, P., \& Scott, P. 2013, Monthly
  Notices of the Royal Astronomical Society, 429, 3658.
\newblock \url{https://doi.org/10.1093/mnras/sts653}

\bibitem[{Sana {et~al.}(2012)Sana, de~Mink, de~Koter, Langer, Evans, Gieles,
  Gosset, Izzard, Le~Bouquin, \& Schneider}]{Sana2012}
Sana, H., de~Mink, S.~E., de~Koter, A., {et~al.} 2012, Science, 337, 444.
\newblock \url{https://science.sciencemag.org/content/337/6093/444}

\bibitem[{{Sander} {et~al.}(2020){Sander}, {Vink}, \&
  {Hamann}}]{2020MNRAS.491.4406S}
{Sander}, A. A.~C., {Vink}, J.~S., \& {Hamann}, W.~R. 2020, \mnras, 491, 4406

\bibitem[{{Saxena} {et~al.}(2020){Saxena}, {Pentericci}, {Mirabelli},
  {Schaerer}, {Schneider}, {Cullen}, {Amorin}, {Bolzonella}, {Bongiorno},
  {Carnall}, {Castellano}, {Cucciati}, {Fontana}, {Fynbo}, {Garilli},
  {Gargiulo}, {Guaita}, {Hathi}, {Hutchison}, {Koekemoer}, {Marchi}, {McLeod},
  {McLure}, {Papovich}, {Pozzetti}, {Talia}, \&
  {Zamorani}}]{2020A&A...636A..47S}
{Saxena}, A., {Pentericci}, L., {Mirabelli}, M., {et~al.} 2020, \aap, 636, A47

\bibitem[{{Schaerer}(2002)}]{Schaerer2002}
{Schaerer}. 2002, A\&A, 382, 28.
\newblock \url{https://doi.org/10.1051/0004-6361:20011619}

\bibitem[{{Schaerer} {et~al.}(2019){Schaerer}, {Fragos}, \&
  {Izotov}}]{2019A&A...622L..10S}
{Schaerer}, D., {Fragos}, T., \& {Izotov}, Y.~I. 2019, \aap, 622, L10

\bibitem[{{Schootemeijer} \& {Langer}(2018)}]{2018A&A...611A..75S}
{Schootemeijer}, A., \& {Langer}, N. 2018, \aap, 611, A75

\bibitem[{{Secunda} {et~al.}(2020){Secunda}, {Cen}, {Kimm}, {G{\"o}tberg}, \&
  {de Mink}}]{Secunda2020}
{Secunda}, A., {Cen}, R., {Kimm}, T., {G{\"o}tberg}, Y., \& {de Mink}, S.~E.
  2020, \apj, 901, 72

\bibitem[{{Senchyna} {et~al.}(2020){Senchyna}, {Stark}, {Mirocha}, {Reines},
  {Charlot}, {Jones}, \& {Mulchaey}}]{2020MNRAS.494..941S}
{Senchyna}, P., {Stark}, D.~P., {Mirocha}, J., {et~al.} 2020, \mnras, 494, 941

\bibitem[{{Shenar} {et~al.}(2019){Shenar}, {Sablowski}, {Hainich}, {Todt},
  {Moffat}, {Oskinova}, {Ramachandran}, {Sana}, {Sander}, {Schnurr},
  {St-Louis}, {Vanbeveren}, {G{\"o}tberg}, \& {Hamann}}]{2019A&A...627A.151S}
{Shenar}, T., {Sablowski}, D.~P., {Hainich}, R., {et~al.} 2019, \aap, 627, A151

\bibitem[{{Shivaei} {et~al.}(2018){Shivaei}, {Reddy}, {Siana}, {Shapley},
  {Kriek}, {Mobasher}, {Freeman}, {Sanders}, {Coil}, {Price}, {Fetherolf},
  {Azadi}, {Leung}, \& {Zick}}]{Shivaei2018}
{Shivaei}, I., {Reddy}, N.~A., {Siana}, B., {et~al.} 2018, \apj, 855, 42

\bibitem[{{Smit} {et~al.}(2016){Smit}, {Bouwens}, {Labb{\'e}}, {Franx},
  {Wilkins}, \& {Oesch}}]{2016ApJ...833..254S}
{Smit}, R., {Bouwens}, R.~J., {Labb{\'e}}, I., {et~al.} 2016, \apj, 833, 254

\bibitem[{{Smith} {et~al.}(2002){Smith}, {Norris}, \&
  {Crowther}}]{2002MNRAS.337.1309S}
{Smith}, L.~J., {Norris}, R. P.~F., \& {Crowther}, P.~A. 2002, \mnras, 337,
  1309

\bibitem[{{Smith}(2014)}]{2014ARA&A..52..487S}
{Smith}, N. 2014, \araa, 52, 487

\bibitem[{Stacy {et~al.}(2016)Stacy, Bromm, \& Lee}]{Stacy2016}
Stacy, A., Bromm, V., \& Lee, A.~T. 2016, Monthly Notices of the Royal
  Astronomical Society, 462, 1307.
\newblock \url{https://doi.org/10.1093/mnras/stw1728}

\bibitem[{{Stanway} {et~al.}(2016){Stanway}, {Eldridge}, \&
  {Becker}}]{2016MNRAS.456..485S}
{Stanway}, E.~R., {Eldridge}, J.~J., \& {Becker}, G.~D. 2016, \mnras, 456, 485

\bibitem[{{Stanway} {et~al.}(2014){Stanway}, {Eldridge}, {Greis}, {Davies},
  {Wilkins}, \& {Bremer}}]{2014MNRAS.444.3466S}
{Stanway}, E.~R., {Eldridge}, J.~J., {Greis}, S. M.~L., {et~al.} 2014, \mnras,
  444, 3466

\bibitem[{{Sutherland} \& {Dopita}(1993)}]{Sutherland1993}
{Sutherland}, R.~S., \& {Dopita}, M.~A. 1993, \apjs, 88, 253

\bibitem[{{Sz{\'e}csi} {et~al.}(2015){Sz{\'e}csi}, {Langer}, {Yoon}, {Sanyal},
  {de Mink}, {Evans}, \& {Dermine}}]{2015A&A...581A..15S}
{Sz{\'e}csi}, D., {Langer}, N., {Yoon}, S.-C., {et~al.} 2015, \aap, 581, A15

\bibitem[{{Tan} \& {McKee}(2004)}]{Tan2004}
{Tan}, J.~C., \& {McKee}, C.~F. 2004, \apj, 603, 383

\bibitem[{{Teyssier}(2002)}]{Teyssier2002}
{Teyssier}. 2002, A\&A, 385, 337.
\newblock \url{https://doi.org/10.1051/0004-6361:20011817}

\bibitem[{Trac {et~al.}(2015)Trac, Cen, \& Mansfield}]{Trac_2015}
Trac, H., Cen, R., \& Mansfield, P. 2015, The Astrophysical Journal, 813, 54.
\newblock \url{https://doi.org/10.1088/0004-637x/813/1/54}

\bibitem[{Tumlinson \& Shull(2000)}]{Tumlinson_2000}
Tumlinson, \& Shull. 2000, The Astrophysical journal, 528 2, L65

\bibitem[{{Van Bever} {et~al.}(1999){Van Bever}, Belkus, Vanbeveren, \& {Van
  Rensbergen}}]{VanBever1999}
{Van Bever}, J., Belkus, H., Vanbeveren, D., \& {Van Rensbergen}, W. 1999, New
  Astronomy, 4, 173 .
\newblock
  \url{http://www.sciencedirect.com/science/article/pii/S1384107699000111}

\bibitem[{{Vanzella} {et~al.}(2018){Vanzella}, {Nonino}, {Cupani},
  {Castellano}, {Sani}, {Mignoli}, {Calura}, {Meneghetti}, {Gilli}, {Comastri},
  {Mercurio}, {Caminha}, {Caputi}, {Rosati}, {Grillo}, {Cristiani}, {Balestra},
  {Fontana}, \& {Giavalisco}}]{2018MNRAS.476L..15V}
{Vanzella}, E., {Nonino}, M., {Cupani}, G., {et~al.} 2018, \mnras, 476, L15

\bibitem[{{Wang} {et~al.}(2018){Wang}, {Gies}, \& {Peters}}]{Wang2018}
{Wang}, L., {Gies}, D.~R., \& {Peters}, G.~J. 2018, \apj, 853, 156

\bibitem[{Weingartner \& Draine(2001)}]{Weingartner_2001}
Weingartner, J.~C., \& Draine, B.~T. 2001, The Astrophysical Journal, 548, 296

\bibitem[{{White} \& {Rees}(1978)}]{1978MNRAS.183..341W}
{White}, S.~D.~M., \& {Rees}, M.~J. 1978, \mnras, 183, 341

\bibitem[{{Wise} \& {Cen}(2009)}]{2009Wise}
{Wise}, J.~H., \& {Cen}, R. 2009, \apj, 693, 984

\bibitem[{{Woods} \& {Gilfanov}(2013)}]{2013MNRAS.432.1640W}
{Woods}, T.~E., \& {Gilfanov}, M. 2013, \mnras, 432, 1640

\bibitem[{{Xiao} {et~al.}(2018){Xiao}, {Stanway}, \&
  {Eldridge}}]{2018MNRAS.477..904X}
{Xiao}, L., {Stanway}, E.~R., \& {Eldridge}, J.~J. 2018, \mnras, 477, 904

\bibitem[{{Yoon} \& {Langer}(2005)}]{2005A&A...443..643Y}
{Yoon}, S.~C., \& {Langer}, N. 2005, \aap, 443, 643

\bibitem[{Zackrisson {et~al.}(2011)Zackrisson, Rydberg, Schaerer, Östlin, \&
  Tuli}]{Zackrisson_2011}
Zackrisson, E., Rydberg, C.-E., Schaerer, D., Östlin, G., \& Tuli, M. 2011,
  The Astrophysical Journal, 740, 13.
\newblock \url{https://doi.org/10.1088%2F0004-637x%2F740%2F1%2F13}

\end{thebibliography}

\end{document}